\documentclass[%
    aip,
    jcp,
    amsmath,
    amssymb,
    twocolumn,
    superscriptaddress,
    reprint]{revtex4-1}

\usepackage{graphicx}%
\usepackage{dcolumn}
\usepackage{amsmath,mathtools}   
\usepackage[usenames,dvipsnames]{pstricks}
\usepackage{pst-grad} 
\usepackage{pst-plot} 
\usepackage[small,compact]{titlesec} 
\usepackage{bm}
\usepackage{color}

\begin{document}

\title{Screening of Coulomb interactions in liquid dielectrics}
\author{Salman Seyedi}
\author{Daniel R.\ Martin}
\affiliation{Department of Physics, Arizona State University,  PO Box 871504, Tempe, AZ 85287-1504 }
\author{Dmitry V.\ Matyushov$^*$ }
\affiliation{Department of Physics and School of Molecular Sciences, Arizona State University,  PO Box 871504, Tempe, AZ 85287-1504 }
\email{dmitrym@asu.edu}

\begin{abstract}
 The interaction of charges in dielectric materials is screened by the dielectric constant of the bulk dielectric. In dielectric theories, screening is assigned to the surface charge appearing from preferential orientations of dipoles along the local field in the interface. For liquid dielectrics, such interfacial orientations are affected by the interfacial structure characterized by a separate interfacial dielectric susceptibility. We argue that dielectric properties of polar liquids should be characterized by two distinct susceptibilities responsible for local response (solvation) and long-range response (dielectric screening). We develop a microscopic model of screening showing that the standard bulk dielectric constant is responsible for screening at large distances. The potential of mean force between ions in polar liquids becomes oscillatory at short distances. Oscillations arise from the coupling of the collective longitudinal excitations of the dipoles in the bulk with the interfacial structure of the liquid around the solutes.           
\end{abstract}
\maketitle

\section{Introduction}
\label{sec:1}
The material formulation of the Coulomb law suggests that the potential energy of two charges, $q_1$ and $q_2$, placed in a dielectric material with the dielectric constant $\epsilon$ should be determined from the equation
\begin{equation}
U = \frac{q_1q_2}{\epsilon R}   . 
\label{eq1}
\end{equation}
The dielectric is then said to screen the interaction between two charges placed at distance $R$, lowering the interaction energy from its vacuum value $q_1q_2/R$ to a value $\epsilon$ times smaller. While the language of interaction energy is often used in electrostatics, $U$ is in fact a potential of mean force (PMF), a free energy, as is now well understood\cite{Huston:1989is,Rashin:1989aa,Bader:1992hm,Fennell:2009fe,Luo:2013dl,Trzesniak:2007aa} and will also become clear from the discussion presented below.  

\begin{figure}
\includegraphics*[width=4cm]{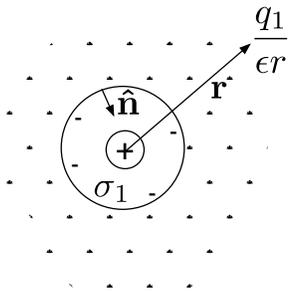}
\caption{Schematic representation of screening of charge $q_1$ by the dielectric with the dielectric constant $\epsilon$. Surface charge with the charge density $\sigma_1=P_n=\mathbf{P}\cdot\mathbf{\hat n}$ develops at the dividing surface. The electrostatic potentials produced by the charge $q_1$ and by the oppositely charged surface charge density $\sigma_1$ add up to $q_1/(\epsilon r)$ inside the dielectric. }
\label{fig:1}  
\end{figure}

Dielectric screening is assigned in theories of dielectrics to the surface charge created at the dividing dielectric surface. For instance, when an ion with the charge $q$ is introduced in the dielectric, the surface charge of an opposite sign is placed at the cavity expelled by the ion from the dielectric material (Fig.\ \ref{fig:1}). Maxwell thought of the surface charge as the result of deformation of the entire material made of positively and negatively charged fluids neutralizing each other.\cite{Maxwell:V1} The external field then deforms the material by pulling and pushing the oppositely charged liquid in opposite directions and creating opposite charges at the dividing surfaces. This view might still apply to an ionic crystal, but needs revision when molecular polar materials are concerned. The current view of interfacial dielectric polarization of polar molecular materials is that molecular dipoles are oriented by the field and predominantly point their oppositely charged ends toward the external charges. Even though they move randomly by thermal agitation, a larger fraction of molecules arrives at the interface oriented along the field thus producing an overall surface charge density of the sign opposite to the sign of the external charge.\cite{Jackson:99} 

The mathematics built around this picture assigns the surface charge density $\sigma$ to the projection of the polarization density $\mathbf{P}$ of the material onto the unit vector $\mathbf{\hat n}$ normal to the dividing surface and pointing outward from the dielectric:\cite{Landau8} : $\sigma = P_n=\mathbf{\hat n}\cdot\mathbf{P}$ (Fig.\ \ref{fig:1}). The surface charge density at the cavity surrounding the charge $q_1$ is then $\sigma_1 = -(q_1/S)(1-\epsilon^{-1})$, where $S=4\pi a^2$ is the surface area of the cavity with the radius $a$. The electrostatic potential of charge $q_1$ and the potential of the opposite charge distributed over the cavity surface add up to $\phi_1=q_1/(\epsilon r)$ at any $r>a$. This electrostatic potential then interacts with the charge $q_2$ with the energy $U=q_2\phi_1$, thus recovering Eq.\ \eqref{eq1}. Importantly, the electrostatic potential in the medium is a small number produced by a nearly complete compensation of two large numbers of opposite sign: the vacuum potential and the potential of the surface charge. This mathematics puts a significant constraint on the accuracy of theoretical formalisms, which should incorporate this compensatory effect before any approximations have been introduced.      

This textbook consideration, and more elaborate derivations,\cite{DMjcp3:14} make a case for a proposal that screening of charges in the bulk of a dipolar dielectric is a surface phenomenon dictated by the orientational structure of dipoles in the interface. If this assumption is correct, then the statistics of material's dipoles pointing their opposite ends to the external charge cannot be determined solely by bulk properties of the material and should be a function of the interfacial structure as well. While Maxwell's notion of bulk deformation still applies to ionic lattices, the focus on the interface seems to be particularly important for liquid dielectrics which respond to inserting a solute by altering their interfacial structure, both in terms of dipolar orientations and interfacial density. The goal of this article is to investigate physical consequences of this proposition and to develop a mathematical formalism to correct Eq.\ \eqref{eq1}. Our focus here is on liquid dielectrics which, according to the picture of interfacial polarization, can build global dielectric screening through changes in the interfacial structure. We show below that, in agreement with standard expectations, the bulk dielectric constant and not interfacial structure ultimately determine the long-distance screening of charges. The interfacial structure affects screening at short distances only.  

The fact that the surface charge density can be significantly modified in polar liquids compared to the standard prescriptions of dielectric theories can be established by numerical simulations of microscopic interfaces. One needs, in accord with the standard rules, the statistical average of the normal projection of the polarization density $\langle P_n\rangle$. It can be calculated from the fluctuation relation\cite{DMjcp3:14,DMjcp3:16} 
\begin{equation}
\langle P_n\rangle = -\beta \langle \delta P_n \delta U^C\rangle, 
\label{eq2} 
\end{equation}
where $\delta U^C$ is the fluctuation of the Coulomb interaction between the charge and the polar medium and $\beta=(k_\text{B}T)^{-1}$ is the inverse temperature. This fluctuation formula was indeed evaluated from molecular dynamics (MD) trajectories obtained for a model nonpolar Kihara solute and corresponding solutes carrying ionic charges.\cite{DMjcp3:16} The result of this calculation was the effective dielectric constant of the interface $\epsilon_\text{int}\simeq 9$ for $a=5$ \AA. The result is obviously much lower that the dielectric constant of bulk water (TIP3P water with $\epsilon \simeq 97$ in the simulations). A low value of an effective dielectric constant around $\epsilon\simeq 5$ has long been suggested to explain ionic mobility\cite{Stiles:1982ck} and in fact was successfully used to calculate ionic activity coefficients.\cite{Lenart:2007df} We stress that essentially equal interface dielectric constants were found for both neutral and ionic solutes with $a=5$ \AA,\cite{DMjcp3:16} suggesting that $\epsilon_\text{int}$ lower than the bulk value can potentially apply not only to ions.   

The question we address here is what is the dielectric constant that should be used in Eq.\ \eqref{eq1} at distances of the nanometer scale. We develop an analytical theory of microscopic screening by a polar liquid and perform molecular dynamics simulations of model solutes in SPC/E water. The main result of the proposed theory is the fluctuation relation for the screening between the charges in the dielectric and the perturbation theory formulated in terms of microscopic pair correlation functions. It casts the screening of charges by a polar liquid in terms of the structure factor of the longitudinal collective excitations of the liquid dipoles. The exact result of this consideration is the following relation
\begin{equation}
U(R) =   \frac{q_1q_2}{\epsilon R} - q_1q_2 \sum_n I^{(n)}(R) . 
\label{eq42}
\end{equation}
Here, the first term is the standard dielectric result in Eq.\ \eqref{eq1}. The second term is the sum over all longitudinal collective excitations of the liquid represented by the poles of the corresponding longitudinal structure factor. These excitations of the bulk liquid dielectric are coupled to the interfacial structure of the solutes to create oscillations of the PMF around the long-distance dielectric result given by Eq.\ \eqref{eq1}. In contrast to screening by free charges in plasmas, where plasmon excitations are quasiparticles with the lifetime significantly exceeding the oscillation period, longitudinal excitations in polar liquids  (dipolarons\cite{Lobo:1973dt,Pollock:1981hc,Madden:84,Omelyan:1996ha,Omelyan:98}) are overdamped.  The  qualitative outcome of the theory is that their overall effect is represented by exponentially decaying oscillations with the decay length $\Lambda$ and the oscillation wavevector $k_\text{max}$ given by the first maximum of the polarization structure factor  
\begin{equation}
\sum_nI^{(n)}(R) \propto e^{-R/\Lambda } \cos \left(k_\text{max} R\right) .
\label{eq20-1}  
\end{equation}

\begin{figure}
  \centering
  \includegraphics*[clip=true,trim= 0cm 1cm 0cm 0cm,width=7.5cm]{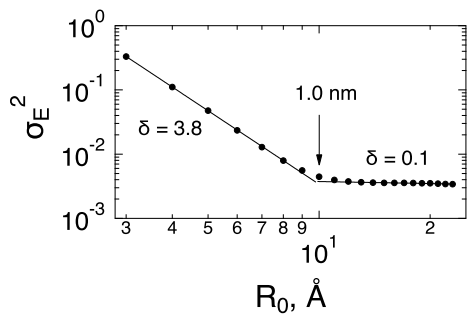}
  \caption{Variance of the electric field of SPC/E water at the
    center of a set of non-polar Kihara solutes with varying size
    $R_{0}$. The solid lines show the fitting of the data with the power law 
    $\sigma_E^2\propto R_{0}^{-\delta}$. The resulting values
    of $\delta$ for smaller and larger solutes are indicated in the plot. The simulations\cite{DMjcp2:11} are done for the Kihara solutes of varying size with the solute-solvent interaction energy $\epsilon_\text{LJ}=0.65$ kJ/mol [Eq.\ \eqref{eq50}]. }
  \label{fig:2}
\end{figure}

Most simulations of the PMF between ions in solution have been performed for small ions typically used as electrolytes.\cite{Gavryushov:2006ku,Fennell:2009fe,Pluharova:2013kt}  The effect of the molecular structure of water on ion pairing is clearly seen in energetic stabilization of contact and solvent-separated ion-pair configurations. The well-defined molecular structure of water around small ions is expected to alter at a nanometer cross-over length-scale,\cite{ChandlerNature:05,Rajamani:05} asymptotically approaching the structure at flat interface. While this cross-over is usually understood in terms of changes in the density profile and shell compressibility,\cite{Sarupria:09} the electrostatic interfacial properties are affected as well.\cite{DMpre1:08,DMjcp2:11} As an example of a dramatic crossover in electrostatic properties, we show in Fig.\ \ref{fig:2} the change of the variance of the solvent field $E_s$ at the center of a spherical solute with the solute size. Applying the linear response approximation, one anticipates that the variance of the solvent field scales as inverse cube\cite{DMpre1:08} of the solute radius $R_0$
\begin{equation}
     \sigma_E^2 =   \beta \sigma^3 \langle (\delta E_{s})^2 \rangle_0
     \propto (\sigma / R_0)^3 ,
     \label{eq20-2} 
\end{equation} 
where the solvent diameter $\sigma$ is used to produce the dimensionless quantity $\sigma_E^2$. For solute radii below $\simeq 1$ nm, the molecular dynamics (MD) simulations\cite{DMjcp2:11} show the power law $\sigma_E^2 \propto R_{0}^{-\delta}$ with $\delta=3.8$ consistent with this expectation. In contrast, there is a sharp cross-over in scaling at $R_{0}\simeq 1$ nm, when the exponent changes to $\delta = 0.1$, i.e., essentially no decay of $\sigma_E^2$ with the growing solute size. These results are reported here based on previously produced trajectories\cite{DMjcp2:11} for Kihara solutes\cite{Kihara:58,DMcpl:11} of varying size. This model solute combines a hard-sphere core with the radius $r_\text{HS}$ with a Lennard-Jones layer of thickness $\sigma_{0s}$ at the surface (see the discussion of the simulation protocol below and, in particular, the interaction potential in Eq.\ \eqref{eq50}). The radii reported in Fig.\ \ref{fig:2}, $R_0=r_\text{HS}+\sigma_{0s}$, are altered by changing $r_\text{HS}$.

The slowing down of the decay of the field variance with increasing solute size is caused by softening of the interface,\cite{ChandlerNature:05} thus allowing stronger fluctuations compensating for an increased size. This crossover does not rule out further crossovers as the size of the solute increases, as we anticipate, but cannot prove with the present computational capabilities. Independently of the long-distance asymptote of $\sigma_E^2$, the appearance of a soft, fluctuating interface raises the question of its coupling with the bulk dipolaron excitations responsible for oscillations in electrostatic screening\cite{Omelyan:98} [Eqs.\ \eqref{eq42} and \eqref{eq20-1}]. For small ions, screening is mostly driven by dielectric laws\cite{Rashin:1989aa} and the structure of the hydration shell is insignificant except at the contact configuration.\cite{Fennell:2009fe} One wonders how extending the size of the solute and changing the density of the hydration layer affect this outcome. Here we report new simulations of the Kihara solutes in SPC/E water\cite{DMjcp2:11} to address these questions. We study how dipolaron excitations in the bulk couple to the interfacial structure and how does this coupling affect oscillations of microscopic dielectric screening. We find that increasing density of the hydration layer, by increasing the solute-solvent attraction, significantly amplifies the PMF oscillations. On the other hand, increasing the size of the solute, beyond the cross-over region in Fig.\ \ref{fig:2}, reduces the oscillations amplitude and leads to a faster approach to the dielectric limit. In other words, softening of the nanometer-scale interface leads to a faster approach to the continuum limit for ionic screening.

\section{Fluctuation relations}
We now consider two charges, $q_1$ and $q_2$, at the distance $R$ immersed in a polar material. Each charge is represented by a repulsive sphere with the radius $a$ defined more precisely below. The total free energy of this system of charges is the sum of their gas-phase interaction energy and the free energy of polarizing the dielectric $F_s$
\begin{equation}
F = \frac{q_1 q_2}{R} + F_s .   
\label{eq3}
\end{equation}
The solvation free energy can be found by thermodynamic integration  in terms of the $\lambda$-scaled Coulomb interaction energy $u_q$ with the solvent\cite{Hansen:13} 
\begin{equation}
F_s=\int_0^1 d\lambda \langle u_q\rangle_\lambda  
\end{equation}
where the average $\langle u_q \rangle_\lambda$ is taken with the statistical ensemble in equilibrium with the solute-solvent potential $\lambda u_q$. One can choose any value $0\le\lambda_0\le 1$ as the reference system and perform a series expansion in terms of the deviation $(\lambda-\lambda_0)u_q$. The result is
\begin{equation}
F_s = \langle u_q\rangle_{\lambda_0} + \left(\lambda_0-\tfrac{1}{2}\right)\beta\langle (\delta u_q)^2\rangle_{\lambda_0} + \dots . 
\label{eq3-1}
\end{equation}
Here, $\delta u_q = u_q - \langle u_q\rangle_{\lambda_0}$ and the interaction energy $u_q$ between the charges and the dielectric is given in terms of the electrostatic potentials $\phi_{si}$ created by the dielectric at the positions of charges $q_i$, $i=1,2$
\begin{equation}
u_q = q_1 \phi_{s1} + q_2 \phi_{s2} .  
\label{eq5}
\end{equation}

In the linear response approximation,\cite{Hummer:1996tv,Aqvist:96}
the infinite expansion  is truncated after the second term in Eq.\ \eqref{eq3-1}. Further, the statistical average $\langle\dots\rangle_{\lambda_0}$ can be performed for the liquid either in equilibrium with $u_q$ and charges $q_i$ or in equilibrium with the repulsive core potentials of the solutes at $q_i=0$, or for any charge state in between.\cite{DMjpca:02} Previous studies\cite{Hummer:1996tv,Aqvist:96,Rajamani:04,DMjcp3:16} have shown that linear response is satisfied exceptionally well when the ionic radius $a$ is sufficiently large to avoid strong interactions between the charges and the dipoles of the medium. We will assume first that this approximation holds and show below that it is indeed satisfied when tested against numerical simulations. In the case of $\lambda_0=0$ we have $q_i=0$ and $\langle u_q\rangle_0=0$. This reference configuration is adopted here for both the analytical derivation and for the numerical simulations discussed below. The solvation free energy $F_s$  is then given by the variance of the interaction energy
\begin{equation}
F_s = - (\beta/2) \langle (\delta u_q)^2\rangle ,  
\label{eq4}
\end{equation}
where we put $\langle\dots \rangle_0 = \langle \dots \rangle$ for brevity. 

The variance of $u_q$ splits into self terms, representing solvation free energies of individual charges, and the cross term modifying their interaction due to the screening by the polar material. Combining the cross terms with the gas-phase interaction energy, we obtain the following formula for the screened interaction energy (PMF) between the charges\cite{Figueirido:1995ti} 
\begin{equation}
  U(R) = q_1q_2\left[R^{-1} - \beta \langle \delta \phi_{s1}\delta \phi_{s2}\rangle \right] .
  \label{eq6} 
\end{equation}
Equation \eqref{eq6} is the starting point for our theoretical development. For the rest of our discussion we put $q_1=q_2=e$, where $e$ is the elementary charge. 

Before we proceed to the formal theory, it is useful to anticipate the result when the standard dielectric theory applies. It is easy to see that Eq.\ \eqref{eq1} is recovered when one assumes for the potential cross correlation
\begin{equation}
  \beta \langle \delta \phi_{s1}\delta \phi_{s2}\rangle = 4\pi \chi^L R^{-1},
  \label{eq6-1}
\end{equation}
where $4\pi\chi^L = 1-\epsilon^{-1}$ is the longitudinal susceptibility of a polar material.\cite{Madden:84} According to the standard expectation of the theory of polar liquids,\cite{Hoye:74} spherical ions interact with the longitudinal polarization of a dipolar liquid with the susceptibility $\chi^L$ in the macroscopic limit of long-wavelength polarization excitations. The theory, therefore, must be able to produce this limit when only the long-ranged macroscopic polarization of the medium is accounted for. The formalism developed next satisfies this expectation.    

\section{Perturbation theory}
If the average $\langle\dots\rangle$ in Eq.\ \eqref{eq6} is treated as an ensemble average over the configurations of a polar liquid around two uncharged cavities with the radii $a$, the calculation of the cross correlation becomes a standard perturbation problem of liquid state theories.\cite{Gubbins:84} One can write the cross correlation in terms of the solute-solvent and solvent-solvent distribution functions as follows  
\begin{equation}\begin{split}
  \langle &\delta \phi_{s}\delta \phi_{s}\rangle = \rho \int d1 \phi_{s}(1)\phi_{s}(1) g_{0s}(r_1) \\
  &+ \rho^2 \int d1d2\phi_{s}(1)\phi_{s}(2) g_{0s}(r_1)g_{0s}(r_2)h_{ss}(12) ,
  \end{split}
  \label{eq7}
\end{equation}
where $\rho=N/V$ is the number density of a polar liquid and 
\begin{equation}
\phi_s(1) = - \frac{\mathbf{m}_1\cdot \mathbf{\hat r_1}}{r_1^2}  
\label{eq8}
\end{equation}
is the electrostatic potential of liquid's dipole $\mathbf{m}_1$ at the position $\mathbf{r}_1$ in the liquid, $\mathbf{\hat r}_1=\mathbf{r}_1/r_1$. The positions and orientations of the liquid dipoles are combined into single indexes such as $(1)=(\mathbf{r},\bm{\omega}_1)$ and  $d1=d\mathbf{r}d\bm{\omega}_1/(4\pi)$. We note also that $\langle \phi_s\rangle=0$ when no preferential orientations of liquid's dipoles is anticipated around an uncharged repulsive core of the solute. Further, the Kirkwood superposition approximation\cite{Gubbins:84,Hansen:13} has been applied to the second term in Eq.\ \eqref{eq7} to represent the three-particle solute-solvent-solvent distribution function as the product of the solute-solvent pair distribution function $g_{0s}(r)$ and the solvent-solvent pair correlation function $h_{ss}(12)$. The latter depends on both the distance between two molecules in the liquid $r_{12}$ and their orientations $\bm{\omega}_1$ and $\bm{\omega}_2$.   
 
One can use Fourier transform to re-write Eq.\ \eqref{eq7} in reciprocal space. The transformation to reciprocal space allows one to eliminate the space convolution in the second summand in Eq.\ \eqref{eq7} and present the result in terms of $k$-space structure factors describing collective fluctuations in the liquid. The details of the derivation are given in the supplementary material (SM) and the result of this derivation is the sum of two terms, $I_1(R)$ and $I_2(R)$, representing the corresponding summands in Eq.\ \eqref{eq7} as one-dimensional $k$-integrals 
\begin{equation}
  I(R) = \beta \langle \delta \phi_{s1} \delta \phi_{s2}\rangle = I_1(R) + I_2(R), 
  \label{eq8-1}
\end{equation}
where
\begin{equation}
\begin{split}
  I_1(R)  & =  \frac{6y}{\pi} \int_0^\infty dk f_{0s}(k) j_0(kR),\\
  I_2(R)  & =  \frac{6y}{\pi}\int_0^{\infty} dk f_{0s}(k)^2 j_0(kR) \left[S^L(k)-1\right] .
  \end{split} 
  \label{eq9}
\end{equation}
Here, $y=(4\pi/9)\beta\rho m^2$ is the standard polarity parameter of the theory of polar liquids,\cite{Gubbins:84,SPH:81} $j_n(x)$ is the spherical Bessel function of $n$th order,\cite{Abramowitz:72} and $f_{0s}(k)$ appears as a result of Fourier transforming $\phi_s(1) g_{0s}(r_1)$. It is given by the relation
\begin{equation}
f_{0s}(k)= k\int_0^\infty dr j_1(kr) g_{0s}(r) , 
\label{eq10}
\end{equation}
which is a special case of the Hankel transform.\cite{SPH:81} Further, $S^L(k)$ in Eq.\ \eqref{eq9} is the longitudinal structure factor of a polar liquid,\cite{Madden:84,Raineri:92} which describes correlated fluctuation of the reciprocal-space polarization density projected on the direction of the wavevector $\mathbf{\hat k}=\mathbf{k}/k$
\begin{equation}
\tilde P^L(\mathbf{k}) = \sum_j (\mathbf{m}_j\cdot \mathbf{\hat k})\ e^{i\mathbf{k}\cdot\mathbf{r}_j}  ,
\label{eq11}
\end{equation}
where the sum runs over all dipoles $\mathbf{m}_j$ in the liquid with their positions at $\mathbf{r}_j$. The structure factor is a scaled variance of this collective variable given as
\begin{equation}
S^L(k) = \frac{3}{Nm^2} \langle \tilde P^L(\mathbf{k})\tilde P^L(-\mathbf{k})\rangle .  
\label{eq12}  
\end{equation}

\begin{figure}
 \includegraphics*[clip=true,trim= 0cm 1cm 0cm 0cm,width=8.5cm]{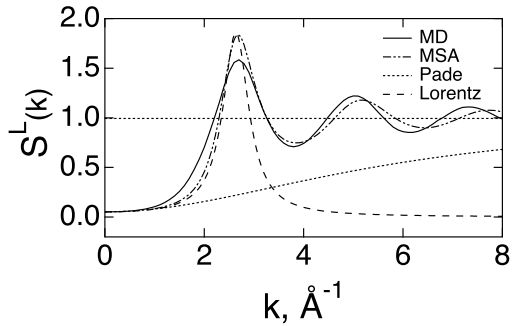}
\caption{$S^L(k)$ for SPC/E water at $T=300$ K from molecular dynamics simulations (MD) and from the MSA solution for dipolar hard spheres\cite{Wertheim:71} in Eq.\ \eqref{eq21} (MSA). The dotted line refers to the Pad{\'e} form in Eq.\ \eqref{eq19} ($\Lambda=0.17$ \AA) and the dashed line marks the Lorentz approximation [Eq.\ \eqref{eq19-1}]. The horizontal dotted line marks the $k\to\infty$ limit $S^L(\infty)\to 1$. }
\label{fig:3} 
\end{figure}

The long-wavelength limit of the structure factor is related to the longitudinal susceptibility of a dielectric through the dimensionless density of dipoles in the liquid $y=(4\pi/9)\beta m^2\rho$ by the following relation
\begin{equation}
3yS^L(0) = 4\pi\chi^L .  
\label{eq13} 
\end{equation}
The opposite limit of $S^L(k)$ at $k\to\infty$ corresponds to disappearance of correlations between different dipoles in the liquid, which leads to $S^L(\infty)=1$. Both limits are illustrated in Fig.\ \ref{fig:3} for SPC/E water from our simulations discussed in more detail below. 

Equations \eqref{eq8-1} and \eqref{eq9} is a formally exact solution for the electrostatic potential cross-correlation within the limits of the linear response approximation and the Kirkwood anzatz\cite{Hansen:13} for the triple solute-solvent-solvent correlation function. We also neglect all multipoles higher than the dipole from the response of the liquid. Our simulations below show that this approximation is adequate for water in contact with relatively large solutes considered here.  

The first summand in Eq.\ \eqref{eq8-1}, $I_1(R)$, describes fluctuations of the potential at two ions produced by rotations and translations of a single molecule in the liquid. The second term, $I_2(R)$, corresponds to correlated thermal motions of two molecules. The interaction of a liquid dipole with the first charge is propagated through the liquid dipole-dipole correlations to the second dipole and then to the second charge. The function $f_{0s}(k)$ reflects the local structure of the liquid around each solute thus coupling the screening excitations of the bulk with the interfacial structure. It has an important property of $f_{0s}(0)=1$  (see below) and scales at large $k$ as $\exp[i k a]$. This latter property allows one to convert $I_i(R)$ into the residue integrals in the complex $k$-space. The integral $I_1(R)$ is calculated exactly as $I_1(R)=3y/R$ if only the pole at $k=0$ is accounted for. The same applies to the $k=0$ pole of $I_2(R)$. Given that we assume $R>2a$, the $k=0$ pole produces the result $I_2^{(0)}(R)=(3y/R)(S^L(0)-1)$. The rest of the contour integral in the complex $k$-space is given by residues of $S^L(k)-1$ at complex poles $k_n$. The final result is 
\begin{equation}
  I(R) = R^{-1}\left(1-\epsilon^{-1}\right) + \sum_n I^{(n)}(R) . 
  \label{eq41}
\end{equation}     
This form can be substituted to Eq.\ \eqref{eq6} with the result for the interaction of two ions given by Eq.\ \eqref{eq42} where also the physical meaning of the microscopic screening terms $I^{(n)}(R)$ is discussed [Eq.\ \eqref{eq20-1}]. 

A significant advantage of the result in Eq.\ \eqref{eq42} is that it incorporates the cancellation of two large terms from the gas-phase (vacuum) Coulomb interaction of the charges and its screening by surface charges of the dielectric as the zero-order term, thus avoiding errors from incorporating approximations into each of the components. The corrections to the continuum limit arise from longitudinal collective excitations in the polar liquid coupled to the interfacial structure of each solute. This is a physically attractive picture, which might extend beyond the derivation presented here. We explore physical consequences of it  in terms of an analytical solution when the poles of the longitudinal structure factor can be well defined.  

Before we turn to this next step, it is useful to identify the approximations made in deriving Eq.\ \eqref{eq41}. First, we have assumed that there is no specific orientational structure of the solvent dipoles around a nonpolar solute carrying zero charge. This is a reasonable approximation in most cases, although water dipoles attend preferential orientations around nonpolar solutes.\cite{Lee:84} This pattern, also found for SPC/E water employed here, tends to diminish when more accurate force fields are used.\cite{Remsing:2014fo} Second, the structure factor $S^L(k)$ in Eq.\ \eqref{eq9} refers to the reference system, which is the polar liquid with inserted nonpolar repulsive cores of the solutes. Since the dielectric constant is affected by solution compared to the bulk, particularly for electrolytes,\cite{Gavryushov:2006ku} the ability to use the structure factor for bulk liquid needs to be tested. We in fact have done this test in our simulations discussed below and have shown that at the concentrations used in our calculations the bulk and solution structure factors are nearly identical (Fig.\ S4 in the SM). We have also tested the sensitivity of the sum over the poles, the second summand in Eq.\ \eqref{eq41}, to the dielectric constant and found it relatively low. The use of the bulk structure factor $S^L(k)$ in Eq.\ \eqref{eq9} is therefore justified and we now proceed to using our analytical approximation to calculate the sum over the dipolaron excitations in the liquid.     
   
\section{Analytical solution}
\begin{figure}
\includegraphics*[clip=true,trim= 0cm 1.5cm 0cm 0cm,width=8.5cm]{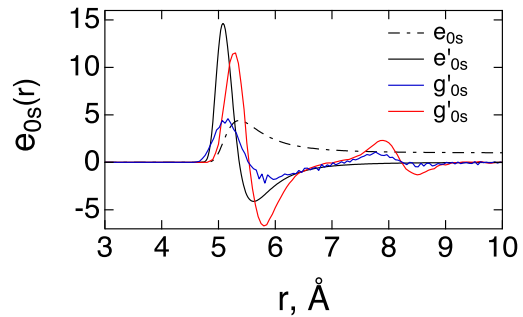}
\caption{Boltzmann factor $e_{0s}(r)$ and its derivative $e_{0s}'(r)$ for the Kihara potential describing the solute-solvent isotropic interaction.  $g_{0s}'(r)$ obtained from molecular dynamics simulations are shown at $\epsilon_\text{LJ}=0.65$ kJ/mol (blue) and 3.7 kJ/mol (red). The position of the positive spike of $e_{0s}'(r)$ defines the cavity radius $a$, which is very close to $R_0=r_\text{HS}+\sigma_{0s}=5$ \AA\ for the Kihara solutes studied here [Eq.\ \eqref{eq50}]. } 
\label{fig:4}  
\end{figure}

In order to study the behavior of $f_{0s}(k)$ in Eq.\ \eqref{eq10}, we will follow here the procedure analogous to that adopted in the perturbation theory of nonpolar (Lennard-Jones) fluids. The theory of nonpolar fluids\cite{WCA76} starts with the observation that the Boltzmann factor, $e(r)=\exp[-\beta u(r)]$, of the intermolecular liquid potential $u(r)$ changes sharply over a short range of distances. This allows one to formulate a perturbation theory in terms of short-ranged ``blip functions''. Following this general framework, we consider the Boltzmann factor of the reference solute-solvent interaction potential $u_{0s}(r)$, which is mostly repulsive and is responsible for the formation of the solute cavity with the radius $a$. The corresponding Boltzmann function,  $e_{0s}(r) = \exp[-\beta u_{0s}(r)]$, of the solute-solvent distance $r$, changes between zero inside the repulsive core of the solute and unity inside the liquid. Figure \ref{fig:4} (dash-dotted line) shows $e_{0s}(r)$ calculated for the solute-water isotropic interaction potential given in the Kihara form\cite{Kihara:58} [Eq.\ \eqref{eq50}] and used in our numerical simulations discussed below. A sharp growth of $e_{0s}(r)$ implies that one can approximate its derivative by a delta-function:\cite{Hansen:13} $e_{0s}'(r)\simeq \delta(r-a)$, which also provides the definition of the cavity radius $a$ as the position of the maximum of $e_{0s}'$.       

We now re-write Eq.\ \eqref{eq10} in the form involving the derivative of the ion-liquid distribution function
\begin{equation}
f_{0s}(k) = \int_0^\infty dr j_0(kr) g_{0s}'(r)  .
\label{eq16} 
\end{equation}
From this equation, one gets at $k=0$ the following boundary condition $f_{0s}(0)=g_{0s}(\infty)-g_{0s}(0)=1$. We next note that $g_{0s}(r)=e_{0s}(r) y_{0s}(r)$, where $y_{0s}(r)$ is a smooth function.\cite{Hansen:13} One therefore can put
\begin{equation}
g_{0s}'(r)\simeq e_{0s}'(r)y_{0s}(r) . 
\label{eq17}
\end{equation}
Figure \ref{fig:4} compares $e_{0s}'(r)$ with $g_{0s}'(r)$ obtained from MD simulations. One can see that $g_{0s}'(r)$ follows the shape of $e_{0s}'(r)$ at the lower value of the solute-solvent Lennard-Jones attraction energy $\epsilon_\text{LJ}$, thus suggesting that $y(r)$ is nearly constant in the range of $r$-values where the spikes of these functions occur. As the attraction increases and the interface becomes more structured, the peak of $g_{0s}'(r)$ shifts to larger distances. Nevertheless, the approximation of $g_{0s}'(r)$ with a positive and negative blips turns out to be quite accurate for modeling $f_{0s}(k)$ at all parameters studied here. 

Given that $e_{0s}'$ involves positive and negative blips (Fig.\ \ref{fig:4}), it can be represented by a sum of two delta-functions positioned at $r=a$ and $r=b$ and carrying positive and negative amplitudes. This transforms $f_{0s}(r)$ to the form  
\begin{equation}
f_{0s}(k) = c j_0(ka) + (1-c) j_0(kb),  
\label{eq18} 
\end{equation}
where $b$ is the position of the negative blip and the coefficients in front of the spherical Bessel functions are chosen to satisfy the condition $f_{0s}(0)=1$. The fit of this function to $f_{0s}(k)$ obtained by numerical integration in Eq.\ \eqref{eq16} is given in the SM. For our discussion here we only need  to know that $f_{0s}(k)^2$ in the integral $I_2(R)$ in Eq.\ \eqref{eq9} scales at most as $e^{\pm 2ibk}$ and can perform the residue integration under the assumption $R>2b$.  

The positions of singularities of $S^L(k)$ in the complex $k$-plane are generally unknown and we resort here to two approximations. We first apply the Ornstein-Zernike approximate based on the known expansion of $S^L(k)$ at low wavevectors.\cite{Hansen:13} According to the Ornstein-Zernike equation, $S^L(k)=\left[1+(\rho/3)c^L(k)\right]^{-1}$, where $c^L(k)$ is the direct correlation function propagating the longitudinal polarization through the liquid.\cite{Wertheim:71,Hansen:13} The expansion $c^L(k)$ in powers of $k$ results in a vanishing linear term such that $S^L(k)^{-1}$ becomes a linear function of $k^2$. One therefore can approximate $S^L(k)$ with the Pad{\'e} form as 
\begin{equation}
S^L(k) = \frac{S^L(0)+\Lambda^2k^2}{1+\Lambda^2 k^2},  
\label{eq19}
\end{equation}
where $\Lambda=0.17$ \AA\ is found from the slope of $S^L(k)^{-1}$ vs $k^2$  for SPC/E water (Fig.\ S5 in SM). This representation of $S^L(k)$ is not very reliable (Fig.\ \ref{fig:3}). A better approximation can be reached in terms of the Lorenzian function with the maximum coinciding with the $k_\text{max}$ of the simulated $S^L(k)$. Since $S^L(k)$ has to be a symmetric function of $k$, $S^L(k)=S^L(-k)$, the following functionality yields a more reasonable approximation (Fig.\ \ref{fig:3})
\begin{equation}
S^L(k) = \tfrac{1}{2} S^L(0)\left[ \frac{k_\text{max}^2+\kappa^2}{(k-k_\text{max})^2+\kappa^2} +  \frac{k_\text{max}^2+\kappa^2}{(k+k_\text{max})^2+\kappa^2} \right] .
\label{eq19-1}  
\end{equation}
This function has two poles in the upper-half $k$-plane: $k_1=k_\text{max} + i\kappa$ and  $k_2=-k_\text{max} + i\kappa$. The sum over these poles results in   
\begin{equation}
\begin{split}
\sum_n I^{(n)}(R) = &\frac{k_\text{max}^2+\kappa^2}{2\kappa R}\left(1-\frac{1}{\epsilon}\right) \\
&\mathrm{Im}\sum_{n=1,2} k_n^{-1} f_{0s}(k_n)^2 e^{iRk_n}.
\end{split}  
\label{eq20}  
\end{equation}

The overdamped dipolar excitations in the polar liquid produce an exponentialy decaying  screening, not unlike the Debye-H{\"u}ckel screening by plasmon excitations in electrolytes [Eq.\ \eqref{eq20-1}]. The fitting of Eq.\ \eqref{eq19-1} to the simulation data produces $k_\text{max}=2.6$ \AA$^{-1}$ and the screening length $\Lambda=\kappa^{-1}=3.2$ \AA, both consistent with the diameter of the water molecule $\sigma\simeq 2.8-2.9$ \AA\ ($2\pi/k_\text{max}=2.4$ \AA). 

The mean-spherical approximation (MSA) for dipolar fluids\cite{Wertheim:71} provides a next step for improving the analytical solution. This exact solution of the Ornstein-Zernike equation with the MSA closure yields the longitudinal structure factor in terms of the Baxter solution\cite{Baxter:70,Hansen:13} $Q(k,\xi^L)$ of the Percus-Yevick closure for the fluid of hard spheres
\begin{equation}
S^L(k)= \left|Q(\kappa^L k,\xi^L)\right|^{-2}  . 
\label{eq21}
\end{equation}
Here, the longitudinal polarity parameters is found from the $k=0$ value of the structure factor  by solving the equation\cite{Wertheim:71}
\begin{equation}
  S^L(0) = \frac{(1-2\xi^L)^4}{(1+4\xi^L)^2} . 
  \label{eq22}
\end{equation}
In addition, an empirical factor $\kappa^L$ is introduced to provide the best fit of the analytical function to the results of simulations. This slight correction is required to reproduce a more open structure of water compared to closely packed simple fluids and results in $\kappa^L=0.85$  for SPC/E water studied here (Fig.\ \ref{fig:3}). Similar scaling is required for other force fields of water when fitted to the Baxter function in Eq.\ \eqref{eq21}. We found $\kappa^L=0.95$\cite{DMcp:06} and $\kappa^L=0.93$\cite{DMjcp1:17} for TIP3P\cite{tip3p:83} and SWM4-DP\cite{Lamoureux:2003cb} water, respectively. The comparison of $S^L(k)$ for the SPC/E and TIP3P water models is shown in Fig.\ S4 in the SM.  

\begin{figure}
 \includegraphics*[clip=true,trim= 0cm 0cm 0cm 0cm,width=7.5cm]{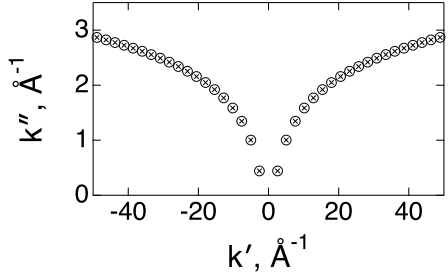}
\caption{Poles $k_n=k_n'+ik_n''$ of the MSA longitudinal structure factor [Eq.\ \eqref{eq21}] in the upper half-plane of the complex $k$-plane: $\left|Q(\kappa^L k_n,\xi^L)\right|^2=0$. The pole closest to the real axis is: $k_1=\pm 2.61 + 0.44 i,$ \AA$^{-1}$. }
\label{fig:5} 
\end{figure}

The analytical form given by Eq.\ \eqref{eq21} results in a large number of poles $k_n=\pm k_{n}' \pm i k_{n}''$ in the complex $k$-plane (Fig.\ \ref{fig:5}). The pole closest to the real axis has its imaginary part corresponding to the correlation length $\Lambda\simeq (k_1'')^{-1} = 2.26$ \AA, reasonably close to the Lorentzian fitting. Further, the pole $I^{(n)}$ in Eq.\ \eqref{eq20} becomes
\begin{equation}
I^{(n)}(R) = \frac{6y}{R} \mathrm{Re} \left[ \frac{f_{0s}(k_n)^2}{k_n c'_n}e^{ik_nR} \right],  
\label{eq23} 
\end{equation}
where $c'_n=(\rho/3)dc^L/dk|_{k=k_n}$.

\begin{figure}
 \includegraphics*[clip=true,trim= 0cm 1cm 0cm 0cm,width=7.5cm]{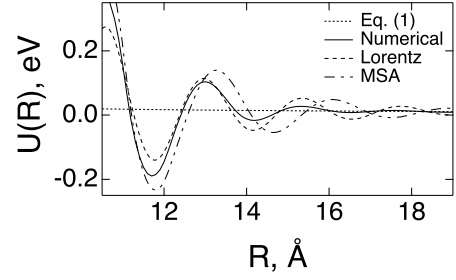}
\caption{Direct integration in Eq.\ \eqref{eq45} (solid line) compared to the Lorentzian approximation in Eq.\ \eqref{eq20} (dashed line) and to the summation over the poles of $S^L(k)$ produced by the MSA (Fig.\ \ref{fig:5}) (dash-dotted line). The calculations are done for two spheres with the radii $5$ \AA\ at varying distance $R$ between their centers. The structure factor for the SPC/E water from simulations (Fig.\ \ref{fig:3}) is used in numerical integration. The corresponding fits to the Lorentz and the MSA solutions are displayed in Fig.\ \ref{fig:3}. The dotted line shows the dielectric result [Eq.\ \eqref{eq1}]. }
\label{fig:6}
\end{figure}

The numerical summation over the poles shown in Fig.\ \ref{fig:5} is compared to both the Lorentz approximation [Eq.\ \eqref{eq20}] and to direct integration (Fig.\ \ref{fig:6}). The latter is done by  combining the integral $I(R)$ in Eq.\ \eqref{eq8-1} and \eqref{eq9} with the Coulomb interaction energy to obtain an integral representation for the PMF
\begin{equation}
U(R) =  \frac{6ye^2}{\pi} \int_0^\infty dk f_{0s}(k)^2 j_0(kR) \left[(3y)^{-1}-S^L(k)\right] . 
\label{eq45}  
\end{equation}
It turns out that the simplest Lorentzian form captures the main features of the PMF, and it is even superior to the summation over the poles produced by the MSA.  The resulting PMF shows oscillations around the continuum solution thus producing over- and under-screening at different distances due to the molecular nature of the polar liquid.\cite{Huston:1989is,Rashin:1989aa} The oscillations of the interaction energy are, however, mostly within $\sim 5-9\ k_\text{B}T$, which is consistent with many previous simulations of ion pairing in force-field water.\cite{Gavryushov:2006ku,Fennell:2009fe,Pluharova:2013kt,DuboueDijon:2018jc}  We now turn to direct MD simulations of the potential cross-correlation in  Eq.\ \eqref{eq6}.

\section{Numerical simulations} 
Numerical MD simulations employed two solutes placed in the simulation box containing 7408 SPC/E\cite{Berendsen:87} water molecules. The solute-solvent interaction potential was given by the isotropic Kihara potential,\cite{Kihara:58,DMcpl:11} which combines the hard-sphere repulsion characterized by the repulsion radius $r_\text{HS}$ with a Lennard-Jones layer of the thickness $\sigma_{0s}$ and the attraction energy $\epsilon_\text{LJ}$ 
\begin{equation} 
u_{0s}(r) =  4 \epsilon_\text{LJ} \left[ \left( \frac{\sigma_{0s}}{r-r_\text{HS}} \right)^{12} - \left( \frac{\sigma_{0s}}{r-r_\text{HS}} \right)^6 \right] \, .
\label{eq50}		
\end{equation}
The Kihara potential was introduced\cite{Kihara:58} to avoid too soft repulsion of the Lennard-Jones potential when applied to sufficiently large solutes.   

The MD trajectories were produced with the NAMD simulation package\cite{Phillips:2005qv} supplemented with a separate script developed to calculate the force between the Kihara solute and SPC/E water. The parameters used for the Kihara potential in this set of simulations were $r_\text{HS}=2$ \AA, $\sigma_{0s}=3$ \AA, and $\epsilon_\text{LJ}=3.7$ kJ/mol. We additionally analyzed the trajectories obtained previously,\cite{DMjcp2:11} which involved the variation of $r_\text{HS}$ to produce the results shown in Fig.\ \ref{fig:2} and for the analysis presented below. We have also analyzed simulation data with changing solute-solvent attraction energy $\epsilon_\text{LJ}$. Two additional values of this parameter, $\epsilon_\text{LJ}=0.65$ kJ/mol and $\epsilon_\text{LJ}=20$ kJ/mol, were used in the analysis. The former attraction energy is close to the interaction energy of the water molecules in the bulk, and it models a hydrophobic solute which does not produce a strong pull on the waters in the hydration shell.\cite{DMcpl:11} The value  $\epsilon_\text{LJ}=3.7$ kJ/mol mostly studied here is more consistent with a hydrophilic solute. Finally, the attraction at $\epsilon_\text{LJ}=20$ kJ/mol is so strong that it breaks water's structure and results in the condensation of the first hydration layer at the solute surface. The resulting layering is seen as a gap of zero value of the solute-solvent pair distribution between the first and second hydration layers (Fig.\ S2 in the SM).      

\begin{figure}
 \includegraphics*[clip=true,trim= 0cm 1cm 0cm 0cm,width=7.5cm]{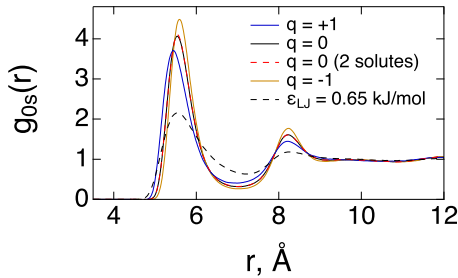}
\caption{Solute-solvent density distribution functions $g_{0s}(r)$ calculated from MD simulations of neutral ($q=0$) and charged ($q=\pm1$) single Kihara solutes in SPC/E water ($r_\text{HS}=2$ \AA). Also shown is the distribution function for a single solute in the box containing two Kihara solutes separated by the distance of $R=20$ \AA. The results shown by the solid lines refer to $\epsilon_\text{LJ}=3.7$ kJ/mol, while the dashed line refers to $\epsilon_\text{LJ}=0.65$ kJ/mol.  }
\label{fig:7}
\end{figure}

One of the advantages of using nonpolar solutes for the calculation of the dipolar screening is that one avoids the Coulomb interactions between the charged solutes and their images in the replicas of the simulation cell, which are unavoidable in any finite-size simulations.\cite{Bader:1992hm} The cross-correlations 
\begin{equation}
I(R)=  \beta \langle \delta \phi_{s1}\delta \phi_{s2}\rangle 
\label{eq51}
\end{equation}
were calculated at a number of configurations with the distance between two Kihara solutes altered in the range $10\le R \le 20$ \AA. However, the ability to use the non-ionic solutes to calculate screening between ions is based on the linear response approximation, which assumes that the solvent structure remains intact for all charge state of the ion, from zero charge to the highest charge considered in this framework. In order to test this assumption we have additionally simulated single Kihara solutes in SPC/E water in neutral and charged states. For the charged solutes, the charge $q=\pm1$ was placed at the center of the Kihara sphere. Figure \ref{fig:7} shows that the pair solute-solvent distribution functions obtained for all three states are very close, in support of the linear response assumption. Further, the solute-solvent density profiles in the simulation box with two solutes are identical to single-solute distribution functions at sufficiently large separations between two spheres (Fig.\ \ref{fig:7}, the two lines are identical on the scale of the plot). The solute-solvent density profile is in fact more strongly affected by the magnitude of the Lennard-Jones energy $\epsilon_\text{LJ}$ in Eq.\ \eqref{eq50} than by the charge state in the range of radii considered here. The dashed line in Fig.\ \ref{fig:7} shows $g_{0s}(r)$ at $\epsilon_\text{LJ}=0.65$ kJ/mol, with a clearly less structured interface.        

\begin{figure}
 \includegraphics*[clip=true,trim= 0cm 0cm 0cm 0cm,width=7.5cm]{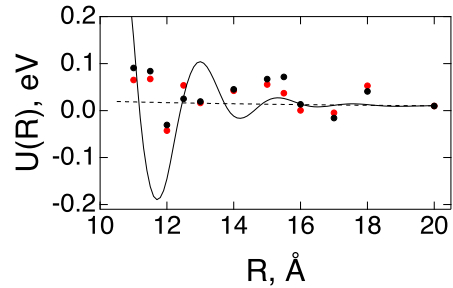}
\caption{Results of MD simulation for two neutral Kihara solutes placed at different distances $R$. Black points refer to electrostatic potential created by water's partial atomic charges and the red points indicate the electrostatic potential created by the water's point dipoles. The solid line is the result of numerical integration in Eq.\  \eqref{eq45} and the dashed line is the dielectric result in Eq.\ \eqref{eq1}.  }
\label{fig:8} 
\end{figure}

The results of calculations of $I(R)$ in Eq.\ \eqref{eq51} need to be combined with the direct Coulomb interaction in Eq.\ \eqref{eq6} to obtain the screened PMF. We found, in agreement with previous results,\cite{Figueirido:1995ti} that this approach leads to the $R\to\infty$ asymptote shifted from zero. The reason is that the Ewald potential $\phi_\text{E}(R)$ is shifted from the Coulomb potential.\cite{Figueirido:1995ti} The simulation results (black points in Fig.\ \ref{fig:8}) were therefore shifted vertically to fit the analytical model (Eq.\ \eqref{eq45}, solid line) at the largest distance studied here. We have additionally performed calculations replacing the atomic charges at the water molecules with point dipoles. These results (red points in Fig.\ \ref{fig:8}) are very close to the charge-based calculations thus justifying the use of the dipolar density field to represent water in the analytical theory. The simulations still do not fit the analytical theory well, which we attribute to the demanding requirement to subtract two nearly-equal quantities to obtaine the PMF.   

\begin{figure}
 \includegraphics*[clip=true,trim= 0cm 1cm 0cm 0cm,width=6.5cm]{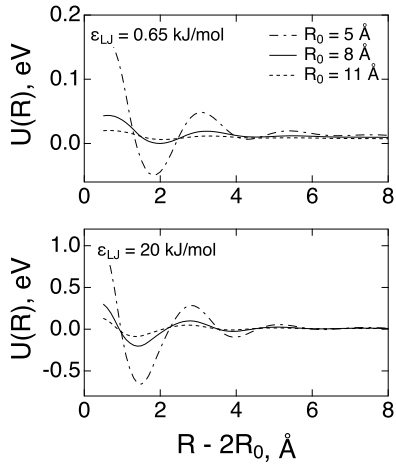}
\caption{$U(R)$ from Eq.\ \eqref{eq45} with $S^L(k)$ for SPC/E water and $f_{0s}(k)$ calculated from solute-water distribution functions of Kihara solutes with changing size $R_0$. The results for two magnitudes of the solute-solvent  Lenard-Jones energy $\epsilon_\text{LJ}$ are shown. }
\label{fig:9}
\end{figure}

We next address the question of the effect of the solute size and the density of the hydration layer on the oscillatory screening behavior of the PMF. Figure \ref{fig:9} shows the calculations performed according to Eq.\ \eqref{eq45} with the solute-water pair distribution functions of Kihara solutes of increasing size. A clear pattern of decreasing amplitude of the screening oscillations is seen for larger solutes. The oscillations essentially disappear beyond the size crossover shown in Fig.\ \ref{fig:2}. The crossover to the nano-scale solvation with soft solute-solvent interface also means the transition to a continuum-type electrostatic screening. 

Increasing the density of the hydration shell produces an opposite effect. We achieve denser hydration layers by significantly increasing the solute-solvent Lennard-Jones attraction (the lower panel in Fig.\ \ref{fig:9}). The value $\epsilon_\text{LJ}=20$ kJ/mol used to illustrate this point is somewhat unrealistic, leading to a collapse of the first hydration shell and layering between the first and second shells (see examples of the solute-solvent distribution function in Fig.\ S2 in the SM). However, this calculation produces an about order-of-magnitude increase in the amplitude of oscillations, indicating that oscillatory pattern of screening is caused by coupling of the bulk dipolarons to the interfacial structure. Increasing the structure of the hydration shell enhances the amplitude of oscillations.

\section{Discussion}
\begin{figure}
 \includegraphics*[clip=true,trim= 0cm 0cm 0cm 0cm,width=4.5cm]{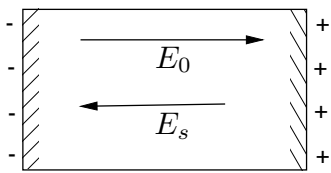}
\caption{Schematics of dielectric screening in solid dielectrics: the external field causes a bulk stress in the sample, resulting in surface charges. The external field $E_0$ is compensated by the field of the surface charges $E_s$ to yield the screened Maxwell field $E=E_0-E_s$. Liquid dielectrics do not support bulk stress and corresponding surface charges must be an interfacial property not directly related to the bulk dielectric constant.  }
\label{fig:10}  
\end{figure}

The textbook picture of screening of electrostatic fields in dielectrics goes back to Maxwell\cite{Maxwell:V1} and considers a slab of dielectric placed in an external field $E_0$. The external field induces bulk strain leading to surface charges. They in turn produce an internal electric field $E_s$ opposing (screening) the external field (Fig.\ \ref{fig:10}). The Maxwell field $E=E_0-E_s$ is the result of near complete cancelation between these two fields, leading to $E_0$ reduced by $\epsilon$. This picture silently assumes that the dielectric is a solid and can sustain bulk stress. The dielectric constant, related to material's ability to develop this bulk stress in response to an external field, is a bulk material property. 

This simple picture is bound to fail and needs to be modified for liquid dielectrics since liquids do not sustain bulk stress and any surface charge must be a surface phenomenon. Since the dielectric constant is still a bulk material property reported by the dielectric experiment, dielectric screening needs to be described in a language disconnected from surface charges. The main question here is whether polarization of the interface and the corresponding interfacial susceptibility, which enters the local polarity response (e.g., for ion solvation), are related to dielectric screening at large (on molecular scale) distances. Not unexpected, our results show that the local response of the liquid interface is mostly unrelated to the long-distance screening. The latter is achieved in liquids by mutual correlations of the liquid dipoles in the bulk and not by the field of the surface charges. The cartoon shown in Fig.\ \ref{fig:10} does not apply to liquid dielectrics, even at the qualitative level. 

A significant consequence of this perspective is that the bulk dielectric constant reported by the dielectric experiment applies to long-distance dielectric screening, but a local interfacial susceptibility has to be used for solvation. In practical terms, polar liquids must be characterized by at least two susceptibilities describing the surface and bulk responses separately. The model solutes dissolved in the force-field water studied here provide a convincing example: their interfacial dielectric constant obtained from Eq.\ \eqref{eq2} is $\simeq 9$,\cite{DMjcp3:16} but the dielectric constant entering the long-distance screening is $\simeq 71$. The analytical theory presented here can be extended to liquids confined in the slab geometry since this extension is achieved at $R_0\to\infty$ while keeping the thickness of the liquid between two solutes constant. The parameters of the theory still remain the same: the density distribution function of the interface and the bulk structure factor.          

We find that the dielectric limit of the Coulomb law in Eq.\ \eqref{eq1} is reached at long distances between the solutes, but the granularity of the polar liquid shows itself over $\simeq 0.5-1$ nm into the bulk in the form of oscillations around the dielectric solution. These oscillations are linked to the overdamped excitations in the polar liquid (dipolarons\cite{Lobo:1973dt,Pollock:1981hc,Madden:84,Omelyan:1996ha,Omelyan:98}) represented by the poles of the longitudinal structure factor of dipolar polarization density.\cite{Pines:99} The excitation with the longest length of decay is responsible for the first peak of the structure factor and is mostly sufficient to reproduce the oscillatory screening calculated by numerical integration. We therefore conclude that dipolaron excitation responsible for the first peak in the structure factor is the main cause of the oscillatory dielectric screening and of the corresponding PMF in ion pairs. It is important to stress that previous reports of oscillatory PMF have been limited to small ions typically employed  as supporting electrolytes.\cite{Rashin:1989aa,Gavryushov:2006ku,Fennell:2009fe,Pluharova:2013kt} Here we show that similar oscillations develop for dielectric screening between large solutes with the diameter of $\simeq 1$ nm.

The simulation protocol employed here is based on the fluctuation relation for the dielectric screening involving the correlation of electrostatic potential produced by the polar liquid at two solutes [Eq.\ \eqref{eq6}]. The advantage of this formalism is that it does not require integrating the force between the solutes over distances.\cite{Pluharova:2013kt,Trzesniak:2007aa}  One therefore can directly calculate the screening between groups belonging to a well-defined structure (such as a protein\cite{DMscirep:17}). Since the approach is based on linear response, one has the freedom to either remove the charges from the corresponding groups or keep them if needed. Nevertheless, dielectric screening is still a challenging task for simulations since subtraction of two large terms is prone to numerical errors. A significant advantage of the theoretical approach summarized in Eq.\ \eqref{eq42} is that subtraction of two largest contributions to the PMF is achieved in the continuum limit and only microscopic corrections linked to damped dipolaron excitations need a separate calculation.

\section*{Supplementary material}
See supplementary material for the simulation protocols and the data analysis.

\acknowledgements 
This research was supported by the Office of Basic Energy Sciences, Division of Chemical Sciences, Geosciences, and Energy Biosciences, Department of Energy (DE-SC0015641). CPU time was provided by the National Science Foundation through XSEDE resources (TG-MCB080071). 

%


\begin{thebibliography}{52}%
\makeatletter
\providecommand \@ifxundefined [1]{%
 \@ifx{#1\undefined}
}%
\providecommand \@ifnum [1]{%
 \ifnum #1\expandafter \@firstoftwo
 \else \expandafter \@secondoftwo
 \fi
}%
\providecommand \@ifx [1]{%
 \ifx #1\expandafter \@firstoftwo
 \else \expandafter \@secondoftwo
 \fi
}%
\providecommand \natexlab [1]{#1}%
\providecommand \enquote  [1]{``#1''}%
\providecommand \bibnamefont  [1]{#1}%
\providecommand \bibfnamefont [1]{#1}%
\providecommand \citenamefont [1]{#1}%
\providecommand \href@noop [0]{\@secondoftwo}%
\providecommand \href [0]{\begingroup \@sanitize@url \@href}%
\providecommand \@href[1]{\@@startlink{#1}\@@href}%
\providecommand \@@href[1]{\endgroup#1\@@endlink}%
\providecommand \@sanitize@url [0]{\catcode `\\12\catcode `\$12\catcode
  `\&12\catcode `\#12\catcode `\^12\catcode `\_12\catcode `\%12\relax}%
\providecommand \@@startlink[1]{}%
\providecommand \@@endlink[0]{}%
\providecommand \url  [0]{\begingroup\@sanitize@url \@url }%
\providecommand \@url [1]{\endgroup\@href {#1}{\urlprefix }}%
\providecommand \urlprefix  [0]{URL }%
\providecommand \Eprint [0]{\href }%
\providecommand \doibase [0]{http://dx.doi.org/}%
\providecommand \selectlanguage [0]{\@gobble}%
\providecommand \bibinfo  [0]{\@secondoftwo}%
\providecommand \bibfield  [0]{\@secondoftwo}%
\providecommand \translation [1]{[#1]}%
\providecommand \BibitemOpen [0]{}%
\providecommand \bibitemStop [0]{}%
\providecommand \bibitemNoStop [0]{.\EOS\space}%
\providecommand \EOS [0]{\spacefactor3000\relax}%
\providecommand \BibitemShut  [1]{\csname bibitem#1\endcsname}%
\let\auto@bib@innerbib\@empty
\bibitem [{\citenamefont {Huston}\ and\ \citenamefont
  {Rossky}(1989)}]{Huston:1989is}%
  \BibitemOpen
  \bibfield  {author} {\bibinfo {author} {\bibfnamefont {S.~E.}\ \bibnamefont
  {Huston}}\ and\ \bibinfo {author} {\bibfnamefont {P.~J.}\ \bibnamefont
  {Rossky}},\ }\href@noop {} {\bibfield  {journal} {\bibinfo  {journal} {J.
  Phys. Chem.}\ }\textbf {\bibinfo {volume} {93}},\ \bibinfo {pages} {7888}
  (\bibinfo {year} {1989})}\BibitemShut {NoStop}%
\bibitem [{\citenamefont {Rashin}(1989)}]{Rashin:1989aa}%
  \BibitemOpen
  \bibfield  {author} {\bibinfo {author} {\bibfnamefont {A.~A.}\ \bibnamefont
  {Rashin}},\ }\href {\doibase 10.1021/j100348a051} {\bibfield  {journal}
  {\bibinfo  {journal} {J. Phys. Chem.}\ }\textbf {\bibinfo {volume} {93}},\
  \bibinfo {pages} {4664} (\bibinfo {year} {1989})}\BibitemShut {NoStop}%
\bibitem [{\citenamefont {Bader}\ and\ \citenamefont
  {Chandler}(1992)}]{Bader:1992hm}%
  \BibitemOpen
  \bibfield  {author} {\bibinfo {author} {\bibfnamefont {J.~S.}\ \bibnamefont
  {Bader}}\ and\ \bibinfo {author} {\bibfnamefont {D.}~\bibnamefont
  {Chandler}},\ }\href@noop {} {\bibfield  {journal} {\bibinfo  {journal} {J.
  Phys. Chem.}\ }\textbf {\bibinfo {volume} {96}},\ \bibinfo {pages} {6423}
  (\bibinfo {year} {1992})}\BibitemShut {NoStop}%
\bibitem [{\citenamefont {Fennell}\ \emph {et~al.}(2009)\citenamefont
  {Fennell}, \citenamefont {Bizjak}, \citenamefont {Vlachy},\ and\
  \citenamefont {Dill}}]{Fennell:2009fe}%
  \BibitemOpen
  \bibfield  {author} {\bibinfo {author} {\bibfnamefont {C.~J.}\ \bibnamefont
  {Fennell}}, \bibinfo {author} {\bibfnamefont {A.}~\bibnamefont {Bizjak}},
  \bibinfo {author} {\bibfnamefont {V.}~\bibnamefont {Vlachy}}, \ and\ \bibinfo
  {author} {\bibfnamefont {K.~A.}\ \bibnamefont {Dill}},\ }\href@noop {}
  {\bibfield  {journal} {\bibinfo  {journal} {J. Phys. Chem. B}\ }\textbf
  {\bibinfo {volume} {113}},\ \bibinfo {pages} {6782} (\bibinfo {year}
  {2009})}\BibitemShut {NoStop}%
\bibitem [{\citenamefont {Luo}\ \emph {et~al.}(2013)\citenamefont {Luo},
  \citenamefont {Jiang}, \citenamefont {Yu}, \citenamefont {MacKerell},\ and\
  \citenamefont {Roux}}]{Luo:2013dl}%
  \BibitemOpen
  \bibfield  {author} {\bibinfo {author} {\bibfnamefont {Y.}~\bibnamefont
  {Luo}}, \bibinfo {author} {\bibfnamefont {W.}~\bibnamefont {Jiang}}, \bibinfo
  {author} {\bibfnamefont {H.}~\bibnamefont {Yu}}, \bibinfo {author}
  {\bibfnamefont {A.~D.}\ \bibnamefont {MacKerell}}, \ and\ \bibinfo {author}
  {\bibfnamefont {B.}~\bibnamefont {Roux}},\ }\href@noop {} {\bibfield
  {journal} {\bibinfo  {journal} {Farad. Disc.}\ }\textbf {\bibinfo {volume}
  {160}},\ \bibinfo {pages} {135} (\bibinfo {year} {2013})}\BibitemShut
  {NoStop}%
\bibitem [{\citenamefont {Trzesniak}, \citenamefont {Kunz},\ and\ \citenamefont
  {van Gunsteren}(2007)}]{Trzesniak:2007aa}%
  \BibitemOpen
  \bibfield  {author} {\bibinfo {author} {\bibfnamefont {D.}~\bibnamefont
  {Trzesniak}}, \bibinfo {author} {\bibfnamefont {A.-P.~E.}\ \bibnamefont
  {Kunz}}, \ and\ \bibinfo {author} {\bibfnamefont {W.~F.}\ \bibnamefont {van
  Gunsteren}},\ }\href {\doibase 10.1002/cphc.200600527} {\bibfield  {journal}
  {\bibinfo  {journal} {ChemPhysChem}\ }\textbf {\bibinfo {volume} {8}},\
  \bibinfo {pages} {162} (\bibinfo {year} {2007})}\BibitemShut {NoStop}%
\bibitem [{\citenamefont {Maxwell}(c 63)}]{Maxwell:V1}%
  \BibitemOpen
  \bibfield  {author} {\bibinfo {author} {\bibfnamefont {J.~C.}\ \bibnamefont
  {Maxwell}},\ }\href@noop {} {\emph {\bibinfo {title} {A Treatise on
  Electricity and Magnetism}}},\ Vol.~\bibinfo {volume} {1}\ (\bibinfo
  {publisher} {Dover Publications},\ \bibinfo {address} {New York},\ \bibinfo
  {year} {1954, sec. 63})\BibitemShut {NoStop}%
\bibitem [{\citenamefont {Jackson}(1999)}]{Jackson:99}%
  \BibitemOpen
  \bibfield  {author} {\bibinfo {author} {\bibfnamefont {J.~D.}\ \bibnamefont
  {Jackson}},\ }\href@noop {} {\emph {\bibinfo {title} {Classical
  {E}lectrodynamics}}}\ (\bibinfo  {publisher} {Wiley},\ \bibinfo {address}
  {New York},\ \bibinfo {year} {1999})\BibitemShut {NoStop}%
\bibitem [{\citenamefont {Landau}\ and\ \citenamefont
  {Lifshitz}(1984)}]{Landau8}%
  \BibitemOpen
  \bibfield  {author} {\bibinfo {author} {\bibfnamefont {L.~D.}\ \bibnamefont
  {Landau}}\ and\ \bibinfo {author} {\bibfnamefont {E.~M.}\ \bibnamefont
  {Lifshitz}},\ }\href@noop {} {\emph {\bibinfo {title} {Electrodynamics of
  {C}ontinuous {M}edia}}}\ (\bibinfo  {publisher} {Pergamon},\ \bibinfo
  {address} {Oxford},\ \bibinfo {year} {1984})\BibitemShut {NoStop}%
\bibitem [{\citenamefont {Matyushov}(2014)}]{DMjcp3:14}%
  \BibitemOpen
  \bibfield  {author} {\bibinfo {author} {\bibfnamefont {D.~V.}\ \bibnamefont
  {Matyushov}},\ }\href@noop {} {\bibfield  {journal} {\bibinfo  {journal} {J.
  Chem. Phys.}\ }\textbf {\bibinfo {volume} {140}},\ \bibinfo {pages} {224506}
  (\bibinfo {year} {2014})}\BibitemShut {NoStop}%
\bibitem [{\citenamefont {Dinpajooh}\ and\ \citenamefont
  {Matyushov}(2016)}]{DMjcp3:16}%
  \BibitemOpen
  \bibfield  {author} {\bibinfo {author} {\bibfnamefont {M.}~\bibnamefont
  {Dinpajooh}}\ and\ \bibinfo {author} {\bibfnamefont {D.~V.}\ \bibnamefont
  {Matyushov}},\ }\href@noop {} {\bibfield  {journal} {\bibinfo  {journal} {J.
  Chem. Phys.}\ }\textbf {\bibinfo {volume} {145}},\ \bibinfo {pages} {014504}
  (\bibinfo {year} {2016})}\BibitemShut {NoStop}%
\bibitem [{\citenamefont {Stiles}, \citenamefont {Hubbard},\ and\ \citenamefont
  {Kayser}(1982)}]{Stiles:1982ck}%
  \BibitemOpen
  \bibfield  {author} {\bibinfo {author} {\bibfnamefont {P.~J.}\ \bibnamefont
  {Stiles}}, \bibinfo {author} {\bibfnamefont {J.~B.}\ \bibnamefont {Hubbard}},
  \ and\ \bibinfo {author} {\bibfnamefont {R.~F.}\ \bibnamefont {Kayser}},\
  }\href@noop {} {\bibfield  {journal} {\bibinfo  {journal} {J. Chem. Phys.}\
  }\textbf {\bibinfo {volume} {77}},\ \bibinfo {pages} {6189} (\bibinfo {year}
  {1982})}\BibitemShut {NoStop}%
\bibitem [{\citenamefont {Lenart}, \citenamefont {Jusufi},\ and\ \citenamefont
  {Panagiotopoulos}(2007)}]{Lenart:2007df}%
  \BibitemOpen
  \bibfield  {author} {\bibinfo {author} {\bibfnamefont {P.~J.}\ \bibnamefont
  {Lenart}}, \bibinfo {author} {\bibfnamefont {A.}~\bibnamefont {Jusufi}}, \
  and\ \bibinfo {author} {\bibfnamefont {A.~Z.}\ \bibnamefont
  {Panagiotopoulos}},\ }\href@noop {} {\bibfield  {journal} {\bibinfo
  {journal} {J. Chem. Phys.}\ }\textbf {\bibinfo {volume} {126}},\ \bibinfo
  {pages} {044509} (\bibinfo {year} {2007})}\BibitemShut {NoStop}%
\bibitem [{\citenamefont {Lobo}, \citenamefont {Robinson},\ and\ \citenamefont
  {Rodriguez}(1973)}]{Lobo:1973dt}%
  \BibitemOpen
  \bibfield  {author} {\bibinfo {author} {\bibfnamefont {R.}~\bibnamefont
  {Lobo}}, \bibinfo {author} {\bibfnamefont {J.~E.}\ \bibnamefont {Robinson}},
  \ and\ \bibinfo {author} {\bibfnamefont {S.}~\bibnamefont {Rodriguez}},\
  }\href@noop {} {\bibfield  {journal} {\bibinfo  {journal} {J. Chem. Phys.}\
  }\textbf {\bibinfo {volume} {59}},\ \bibinfo {pages} {5992} (\bibinfo {year}
  {1973})}\BibitemShut {NoStop}%
\bibitem [{\citenamefont {Pollock}\ and\ \citenamefont
  {Alder}(1981)}]{Pollock:1981hc}%
  \BibitemOpen
  \bibfield  {author} {\bibinfo {author} {\bibfnamefont {E.~L.}\ \bibnamefont
  {Pollock}}\ and\ \bibinfo {author} {\bibfnamefont {B.~J.}\ \bibnamefont
  {Alder}},\ }\href@noop {} {\bibfield  {journal} {\bibinfo  {journal} {Phys.
  Rev. Lett.}\ }\textbf {\bibinfo {volume} {46}},\ \bibinfo {pages} {950}
  (\bibinfo {year} {1981})}\BibitemShut {NoStop}%
\bibitem [{\citenamefont {Madden}\ and\ \citenamefont
  {Kivelson}(1984)}]{Madden:84}%
  \BibitemOpen
  \bibfield  {author} {\bibinfo {author} {\bibfnamefont {P.}~\bibnamefont
  {Madden}}\ and\ \bibinfo {author} {\bibfnamefont {D.}~\bibnamefont
  {Kivelson}},\ }\href@noop {} {\bibfield  {journal} {\bibinfo  {journal} {Adv.
  Chem. Phys.}\ }\textbf {\bibinfo {volume} {56}},\ \bibinfo {pages} {467}
  (\bibinfo {year} {1984})}\BibitemShut {NoStop}%
\bibitem [{\citenamefont {Omelyan}(1996)}]{Omelyan:1996ha}%
  \BibitemOpen
  \bibfield  {author} {\bibinfo {author} {\bibfnamefont {I.~P.}\ \bibnamefont
  {Omelyan}},\ }\href@noop {} {\bibfield  {journal} {\bibinfo  {journal} {Phys.
  Lett. A}\ }\textbf {\bibinfo {volume} {216}},\ \bibinfo {pages} {211}
  (\bibinfo {year} {1996})}\BibitemShut {NoStop}%
\bibitem [{\citenamefont {Omelyan}(1998)}]{Omelyan:98}%
  \BibitemOpen
  \bibfield  {author} {\bibinfo {author} {\bibfnamefont {I.~P.}\ \bibnamefont
  {Omelyan}},\ }\href@noop {} {\bibfield  {journal} {\bibinfo  {journal} {Mol.
  Phys.}\ }\textbf {\bibinfo {volume} {93}},\ \bibinfo {pages} {123} (\bibinfo
  {year} {1998})}\BibitemShut {NoStop}%
\bibitem [{\citenamefont {Gavryushov}(2006)}]{Gavryushov:2006ku}%
  \BibitemOpen
  \bibfield  {author} {\bibinfo {author} {\bibfnamefont {S.}~\bibnamefont
  {Gavryushov}},\ }\href@noop {} {\bibfield  {journal} {\bibinfo  {journal} {J.
  Phys. Chem. B}\ }\textbf {\bibinfo {volume} {110}},\ \bibinfo {pages} {10888}
  (\bibinfo {year} {2006})}\BibitemShut {NoStop}%
\bibitem [{\citenamefont {Pluha{\v{r}}ov{\'a}}\ \emph
  {et~al.}(2013)\citenamefont {Pluha{\v{r}}ov{\'a}}, \citenamefont {Marsalek},
  \citenamefont {Schmidt},\ and\ \citenamefont {Jungwirth}}]{Pluharova:2013kt}%
  \BibitemOpen
  \bibfield  {author} {\bibinfo {author} {\bibfnamefont {E.}~\bibnamefont
  {Pluha{\v{r}}ov{\'a}}}, \bibinfo {author} {\bibfnamefont {O.}~\bibnamefont
  {Marsalek}}, \bibinfo {author} {\bibfnamefont {B.}~\bibnamefont {Schmidt}}, \
  and\ \bibinfo {author} {\bibfnamefont {P.}~\bibnamefont {Jungwirth}},\
  }\href@noop {} {\bibfield  {journal} {\bibinfo  {journal} {J. Phys. Chem.
  Lett.}\ }\textbf {\bibinfo {volume} {4}},\ \bibinfo {pages} {4177} (\bibinfo
  {year} {2013})}\BibitemShut {NoStop}%
\bibitem [{\citenamefont {Chandler}(2005)}]{ChandlerNature:05}%
  \BibitemOpen
  \bibfield  {author} {\bibinfo {author} {\bibfnamefont {D.}~\bibnamefont
  {Chandler}},\ }\href@noop {} {\bibfield  {journal} {\bibinfo  {journal}
  {Nature}\ }\textbf {\bibinfo {volume} {437}},\ \bibinfo {pages} {640}
  (\bibinfo {year} {2005})}\BibitemShut {NoStop}%
\bibitem [{\citenamefont {Rajamani}, \citenamefont {Truskett},\ and\
  \citenamefont {Garde}(2005)}]{Rajamani:05}%
  \BibitemOpen
  \bibfield  {author} {\bibinfo {author} {\bibfnamefont {S.}~\bibnamefont
  {Rajamani}}, \bibinfo {author} {\bibfnamefont {T.~M.}\ \bibnamefont
  {Truskett}}, \ and\ \bibinfo {author} {\bibfnamefont {S.}~\bibnamefont
  {Garde}},\ }\href@noop {} {\bibfield  {journal} {\bibinfo  {journal} {Proc.\
  Natl.\ Acad.\ Sci.}\ }\textbf {\bibinfo {volume} {102}},\ \bibinfo {pages}
  {9475} (\bibinfo {year} {2005})}\BibitemShut {NoStop}%
\bibitem [{\citenamefont {Sarupria}\ and\ \citenamefont
  {Garde}(2009)}]{Sarupria:09}%
  \BibitemOpen
  \bibfield  {author} {\bibinfo {author} {\bibfnamefont {S.}~\bibnamefont
  {Sarupria}}\ and\ \bibinfo {author} {\bibfnamefont {S.}~\bibnamefont
  {Garde}},\ }\href@noop {} {\bibfield  {journal} {\bibinfo  {journal} {Phys.
  Rev. Lett.}\ }\textbf {\bibinfo {volume} {103}},\ \bibinfo {pages} {037803}
  (\bibinfo {year} {2009})}\BibitemShut {NoStop}%
\bibitem [{\citenamefont {Martin}\ and\ \citenamefont
  {Matyushov}(2008)}]{DMpre1:08}%
  \BibitemOpen
  \bibfield  {author} {\bibinfo {author} {\bibfnamefont {D.~R.}\ \bibnamefont
  {Martin}}\ and\ \bibinfo {author} {\bibfnamefont {D.~V.}\ \bibnamefont
  {Matyushov}},\ }\href@noop {} {\bibfield  {journal} {\bibinfo  {journal}
  {Phys. Rev. E}\ }\textbf {\bibinfo {volume} {78}},\ \bibinfo {pages} {041206}
  (\bibinfo {year} {2008})}\BibitemShut {NoStop}%
\bibitem [{\citenamefont {Martin}, \citenamefont {Friesen},\ and\ \citenamefont
  {Matyushov}(2011)}]{DMjcp2:11}%
  \BibitemOpen
  \bibfield  {author} {\bibinfo {author} {\bibfnamefont {D.~R.}\ \bibnamefont
  {Martin}}, \bibinfo {author} {\bibfnamefont {A.~D.}\ \bibnamefont {Friesen}},
  \ and\ \bibinfo {author} {\bibfnamefont {D.~V.}\ \bibnamefont {Matyushov}},\
  }\href@noop {} {\bibfield  {journal} {\bibinfo  {journal} {J. Chem. Phys.}\
  }\textbf {\bibinfo {volume} {135}},\ \bibinfo {pages} {084514} (\bibinfo
  {year} {2011})}\BibitemShut {NoStop}%
\bibitem [{\citenamefont {Kihara}(1958)}]{Kihara:58}%
  \BibitemOpen
  \bibfield  {author} {\bibinfo {author} {\bibfnamefont {T.}~\bibnamefont
  {Kihara}},\ }\href@noop {} {\bibfield  {journal} {\bibinfo  {journal} {Adv.
  Chem. Phys.}\ }\textbf {\bibinfo {volume} {1}},\ \bibinfo {pages} {267}
  (\bibinfo {year} {1958})}\BibitemShut {NoStop}%
\bibitem [{\citenamefont {Friesen}\ and\ \citenamefont
  {Matyushov}(2011)}]{DMcpl:11}%
  \BibitemOpen
  \bibfield  {author} {\bibinfo {author} {\bibfnamefont {A.~D.}\ \bibnamefont
  {Friesen}}\ and\ \bibinfo {author} {\bibfnamefont {D.~V.}\ \bibnamefont
  {Matyushov}},\ }\href@noop {} {\bibfield  {journal} {\bibinfo  {journal}
  {Chem. Phys. Lett.}\ }\textbf {\bibinfo {volume} {511}},\ \bibinfo {pages}
  {256} (\bibinfo {year} {2011})}\BibitemShut {NoStop}%
\bibitem [{\citenamefont {Hansen}\ and\ \citenamefont
  {McDonald}(2013)}]{Hansen:13}%
  \BibitemOpen
  \bibfield  {author} {\bibinfo {author} {\bibfnamefont {J.-P.}\ \bibnamefont
  {Hansen}}\ and\ \bibinfo {author} {\bibfnamefont {I.~R.}\ \bibnamefont
  {McDonald}},\ }\href@noop {} {\emph {\bibinfo {title} {Theory of {S}imple
  {L}iquids}}},\ \bibinfo {edition} {4th}\ ed.\ (\bibinfo  {publisher}
  {Academic Press},\ \bibinfo {address} {Amsterdam},\ \bibinfo {year}
  {2013})\BibitemShut {NoStop}%
\bibitem [{\citenamefont {Hummer}\ and\ \citenamefont
  {Szabo}(1996)}]{Hummer:1996tv}%
  \BibitemOpen
  \bibfield  {author} {\bibinfo {author} {\bibfnamefont {G.}~\bibnamefont
  {Hummer}}\ and\ \bibinfo {author} {\bibfnamefont {A.}~\bibnamefont {Szabo}},\
  }\href@noop {} {\bibfield  {journal} {\bibinfo  {journal} {J. Chem. Phys.}\
  }\textbf {\bibinfo {volume} {105}},\ \bibinfo {pages} {2004} (\bibinfo {year}
  {1996})}\BibitemShut {NoStop}%
\bibitem [{\citenamefont {{\AA}qvist}\ and\ \citenamefont
  {Hansson}(1996)}]{Aqvist:96}%
  \BibitemOpen
  \bibfield  {author} {\bibinfo {author} {\bibfnamefont {J.}~\bibnamefont
  {{\AA}qvist}}\ and\ \bibinfo {author} {\bibfnamefont {T.}~\bibnamefont
  {Hansson}},\ }\href@noop {} {\bibfield  {journal} {\bibinfo  {journal} {J.
  Phys. Chem.}\ }\textbf {\bibinfo {volume} {100}},\ \bibinfo {pages} {9512}
  (\bibinfo {year} {1996})}\BibitemShut {NoStop}%
\bibitem [{\citenamefont {Milischuk}\ and\ \citenamefont
  {Matyushov}(2002)}]{DMjpca:02}%
  \BibitemOpen
  \bibfield  {author} {\bibinfo {author} {\bibfnamefont {A.}~\bibnamefont
  {Milischuk}}\ and\ \bibinfo {author} {\bibfnamefont {D.~V.}\ \bibnamefont
  {Matyushov}},\ }\href@noop {} {\bibfield  {journal} {\bibinfo  {journal} {J.\
  Phys.\ Chem. A}\ }\textbf {\bibinfo {volume} {106}},\ \bibinfo {pages} {2146}
  (\bibinfo {year} {2002})}\BibitemShut {NoStop}%
\bibitem [{\citenamefont {Rajamani}, \citenamefont {Ghosh},\ and\ \citenamefont
  {Garde}(2004)}]{Rajamani:04}%
  \BibitemOpen
  \bibfield  {author} {\bibinfo {author} {\bibfnamefont {S.}~\bibnamefont
  {Rajamani}}, \bibinfo {author} {\bibfnamefont {T.}~\bibnamefont {Ghosh}}, \
  and\ \bibinfo {author} {\bibfnamefont {S.}~\bibnamefont {Garde}},\
  }\href@noop {} {\bibfield  {journal} {\bibinfo  {journal} {J.\ Chem.\ Phys.}\
  }\textbf {\bibinfo {volume} {120}},\ \bibinfo {pages} {4457} (\bibinfo {year}
  {2004})}\BibitemShut {NoStop}%
\bibitem [{\citenamefont {Figueirido}, \citenamefont {Del~Buono},\ and\
  \citenamefont {Levy}(1995)}]{Figueirido:1995ti}%
  \BibitemOpen
  \bibfield  {author} {\bibinfo {author} {\bibfnamefont {F.}~\bibnamefont
  {Figueirido}}, \bibinfo {author} {\bibfnamefont {G.~S.}\ \bibnamefont
  {Del~Buono}}, \ and\ \bibinfo {author} {\bibfnamefont {R.~M.}\ \bibnamefont
  {Levy}},\ }\href@noop {} {\bibfield  {journal} {\bibinfo  {journal} {J. Chem.
  Phys.}\ }\textbf {\bibinfo {volume} {103}},\ \bibinfo {pages} {6133}
  (\bibinfo {year} {1995})}\BibitemShut {NoStop}%
\bibitem [{\citenamefont {H{\o}ye}\ and\ \citenamefont
  {Stell}(1974)}]{Hoye:74}%
  \BibitemOpen
  \bibfield  {author} {\bibinfo {author} {\bibfnamefont {J.~S.}\ \bibnamefont
  {H{\o}ye}}\ and\ \bibinfo {author} {\bibfnamefont {G.}~\bibnamefont
  {Stell}},\ }\href@noop {} {\bibfield  {journal} {\bibinfo  {journal} {J.\
  Chem.\ Phys.}\ }\textbf {\bibinfo {volume} {61}},\ \bibinfo {pages} {562}
  (\bibinfo {year} {1974})}\BibitemShut {NoStop}%
\bibitem [{\citenamefont {Gray}\ and\ \citenamefont
  {Gubbins}(1984)}]{Gubbins:84}%
  \BibitemOpen
  \bibfield  {author} {\bibinfo {author} {\bibfnamefont {C.~G.}\ \bibnamefont
  {Gray}}\ and\ \bibinfo {author} {\bibfnamefont {K.~E.}\ \bibnamefont
  {Gubbins}},\ }\href@noop {} {\emph {\bibinfo {title} {Theory of Molecular
  Liquids}}}\ (\bibinfo  {publisher} {Clarendon Press},\ \bibinfo {address}
  {Oxford},\ \bibinfo {year} {1984})\BibitemShut {NoStop}%
\bibitem [{\citenamefont {Abramowitz}\ and\ \citenamefont
  {Stegun}(1972)}]{Abramowitz:72}%
  \BibitemOpen
  \bibinfo {editor} {\bibfnamefont {M.}~\bibnamefont {Abramowitz}}\ and\
  \bibinfo {editor} {\bibfnamefont {I.~A.}\ \bibnamefont {Stegun}},\ eds.,\
  \href@noop {} {\emph {\bibinfo {title} {Handbook of Mathematical
  Functions}}}\ (\bibinfo  {publisher} {Dover},\ \bibinfo {address} {New
  York},\ \bibinfo {year} {1972})\BibitemShut {NoStop}%
\bibitem [{\citenamefont {Stell}, \citenamefont {Patey},\ and\ \citenamefont
  {H{{\o}}ye}(1981)}]{SPH:81}%
  \BibitemOpen
  \bibfield  {author} {\bibinfo {author} {\bibfnamefont {G.}~\bibnamefont
  {Stell}}, \bibinfo {author} {\bibfnamefont {G.~N.}\ \bibnamefont {Patey}}, \
  and\ \bibinfo {author} {\bibfnamefont {J.~S.}\ \bibnamefont {H{{\o}}ye}},\
  }\href@noop {} {\bibfield  {journal} {\bibinfo  {journal} {Adv. Chem. Phys.}\
  }\textbf {\bibinfo {volume} {48}},\ \bibinfo {pages} {183} (\bibinfo {year}
  {1981})}\BibitemShut {NoStop}%
\bibitem [{\citenamefont {Raineri}, \citenamefont {Resat},\ and\ \citenamefont
  {Friedman}(1992)}]{Raineri:92}%
  \BibitemOpen
  \bibfield  {author} {\bibinfo {author} {\bibfnamefont {F.~O.}\ \bibnamefont
  {Raineri}}, \bibinfo {author} {\bibfnamefont {H.}~\bibnamefont {Resat}}, \
  and\ \bibinfo {author} {\bibfnamefont {H.~L.}\ \bibnamefont {Friedman}},\
  }\href@noop {} {\bibfield  {journal} {\bibinfo  {journal} {J.\ Chem.\ Phys.}\
  }\textbf {\bibinfo {volume} {96}},\ \bibinfo {pages} {3068} (\bibinfo {year}
  {1992})}\BibitemShut {NoStop}%
\bibitem [{\citenamefont {Wertheim}(1971)}]{Wertheim:71}%
  \BibitemOpen
  \bibfield  {author} {\bibinfo {author} {\bibfnamefont {M.~S.}\ \bibnamefont
  {Wertheim}},\ }\href@noop {} {\bibfield  {journal} {\bibinfo  {journal} {J.
  Chem. Phys.}\ }\textbf {\bibinfo {volume} {55}},\ \bibinfo {pages} {4291}
  (\bibinfo {year} {1971})}\BibitemShut {NoStop}%
\bibitem [{\citenamefont {Lee}, \citenamefont {McCammon},\ and\ \citenamefont
  {Rossky}(1984)}]{Lee:84}%
  \BibitemOpen
  \bibfield  {author} {\bibinfo {author} {\bibfnamefont {C.~Y.}\ \bibnamefont
  {Lee}}, \bibinfo {author} {\bibfnamefont {J.~A.}\ \bibnamefont {McCammon}}, \
  and\ \bibinfo {author} {\bibfnamefont {P.~J.}\ \bibnamefont {Rossky}},\
  }\href@noop {} {\bibfield  {journal} {\bibinfo  {journal} {J. Chem. Phys.}\
  }\textbf {\bibinfo {volume} {80}},\ \bibinfo {pages} {4448} (\bibinfo {year}
  {1984})}\BibitemShut {NoStop}%
\bibitem [{\citenamefont {Remsing}\ \emph {et~al.}(2014)\citenamefont
  {Remsing}, \citenamefont {Baer}, \citenamefont {Schenter}, \citenamefont
  {Mundy},\ and\ \citenamefont {Weeks}}]{Remsing:2014fo}%
  \BibitemOpen
  \bibfield  {author} {\bibinfo {author} {\bibfnamefont {R.~C.}\ \bibnamefont
  {Remsing}}, \bibinfo {author} {\bibfnamefont {M.~D.}\ \bibnamefont {Baer}},
  \bibinfo {author} {\bibfnamefont {G.~K.}\ \bibnamefont {Schenter}}, \bibinfo
  {author} {\bibfnamefont {C.~J.}\ \bibnamefont {Mundy}}, \ and\ \bibinfo
  {author} {\bibfnamefont {J.~D.}\ \bibnamefont {Weeks}},\ }\href@noop {}
  {\bibfield  {journal} {\bibinfo  {journal} {J. Phys. Chem. Lett.}\ }\textbf
  {\bibinfo {volume} {5}},\ \bibinfo {pages} {2767} (\bibinfo {year}
  {2014})}\BibitemShut {NoStop}%
\bibitem [{\citenamefont {Andersen}, \citenamefont {Chandler},\ and\
  \citenamefont {Weeks}(1976)}]{WCA76}%
  \BibitemOpen
  \bibfield  {author} {\bibinfo {author} {\bibfnamefont {H.~C.}\ \bibnamefont
  {Andersen}}, \bibinfo {author} {\bibfnamefont {D.}~\bibnamefont {Chandler}},
  \ and\ \bibinfo {author} {\bibfnamefont {J.~D.}\ \bibnamefont {Weeks}},\
  }\href@noop {} {\bibfield  {journal} {\bibinfo  {journal} {Adv. Chem. Phys.}\
  }\textbf {\bibinfo {volume} {34}},\ \bibinfo {pages} {105} (\bibinfo {year}
  {1976})}\BibitemShut {NoStop}%
\bibitem [{\citenamefont {Baxter}(1970)}]{Baxter:70}%
  \BibitemOpen
  \bibfield  {author} {\bibinfo {author} {\bibfnamefont {R.~J.}\ \bibnamefont
  {Baxter}},\ }\href@noop {} {\bibfield  {journal} {\bibinfo  {journal} {J.\
  Chem.\ Phys.}\ }\textbf {\bibinfo {volume} {52}},\ \bibinfo {pages} {4559}
  (\bibinfo {year} {1970})}\BibitemShut {NoStop}%
\bibitem [{\citenamefont {Milischuk}, \citenamefont {Matyushov},\ and\
  \citenamefont {Newton}(2006)}]{DMcp:06}%
  \BibitemOpen
  \bibfield  {author} {\bibinfo {author} {\bibfnamefont {A.~A.}\ \bibnamefont
  {Milischuk}}, \bibinfo {author} {\bibfnamefont {D.~V.}\ \bibnamefont
  {Matyushov}}, \ and\ \bibinfo {author} {\bibfnamefont {M.~D.}\ \bibnamefont
  {Newton}},\ }\href@noop {} {\bibfield  {journal} {\bibinfo  {journal} {Chem.
  Phys.}\ }\textbf {\bibinfo {volume} {324}},\ \bibinfo {pages} {172} (\bibinfo
  {year} {2006})}\BibitemShut {NoStop}%
\bibitem [{\citenamefont {Dinpajooh}, \citenamefont {Newton},\ and\
  \citenamefont {Matyushov}(2017)}]{DMjcp1:17}%
  \BibitemOpen
  \bibfield  {author} {\bibinfo {author} {\bibfnamefont {M.}~\bibnamefont
  {Dinpajooh}}, \bibinfo {author} {\bibfnamefont {M.~D.}\ \bibnamefont
  {Newton}}, \ and\ \bibinfo {author} {\bibfnamefont {D.~V.}\ \bibnamefont
  {Matyushov}},\ }\href@noop {} {\bibfield  {journal} {\bibinfo  {journal} {J.
  Chem. Phys.}\ }\textbf {\bibinfo {volume} {145}},\ \bibinfo {pages} {064504}
  (\bibinfo {year} {2017})}\BibitemShut {NoStop}%
\bibitem [{\citenamefont {Jorgensen}\ \emph {et~al.}(1983)\citenamefont
  {Jorgensen}, \citenamefont {Chandrasekhar}, \citenamefont {Madura},
  \citenamefont {Impey},\ and\ \citenamefont {Klein}}]{tip3p:83}%
  \BibitemOpen
  \bibfield  {author} {\bibinfo {author} {\bibfnamefont {W.~L.}\ \bibnamefont
  {Jorgensen}}, \bibinfo {author} {\bibfnamefont {J.}~\bibnamefont
  {Chandrasekhar}}, \bibinfo {author} {\bibfnamefont {J.~D.}\ \bibnamefont
  {Madura}}, \bibinfo {author} {\bibfnamefont {R.~W.}\ \bibnamefont {Impey}}, \
  and\ \bibinfo {author} {\bibfnamefont {M.~L.}\ \bibnamefont {Klein}},\
  }\href@noop {} {\bibfield  {journal} {\bibinfo  {journal} {J. Chem. Phys.}\
  }\textbf {\bibinfo {volume} {79}},\ \bibinfo {pages} {926} (\bibinfo {year}
  {1983})}\BibitemShut {NoStop}%
\bibitem [{\citenamefont {Lamoureux}\ and\ \citenamefont
  {Roux}(2003)}]{Lamoureux:2003cb}%
  \BibitemOpen
  \bibfield  {author} {\bibinfo {author} {\bibfnamefont {G.}~\bibnamefont
  {Lamoureux}}\ and\ \bibinfo {author} {\bibfnamefont {B.}~\bibnamefont
  {Roux}},\ }\href@noop {} {\bibfield  {journal} {\bibinfo  {journal} {J. Chem.
  Phys.}\ }\textbf {\bibinfo {volume} {119}},\ \bibinfo {pages} {3025}
  (\bibinfo {year} {2003})}\BibitemShut {NoStop}%
\bibitem [{\citenamefont {Dubou{\'e}-Dijon}\ \emph {et~al.}(2018)\citenamefont
  {Dubou{\'e}-Dijon}, \citenamefont {Mason}, \citenamefont {Fischer},\ and\
  \citenamefont {Jungwirth}}]{DuboueDijon:2018jc}%
  \BibitemOpen
  \bibfield  {author} {\bibinfo {author} {\bibfnamefont {E.}~\bibnamefont
  {Dubou{\'e}-Dijon}}, \bibinfo {author} {\bibfnamefont {P.~E.}\ \bibnamefont
  {Mason}}, \bibinfo {author} {\bibfnamefont {H.~E.}\ \bibnamefont {Fischer}},
  \ and\ \bibinfo {author} {\bibfnamefont {P.}~\bibnamefont {Jungwirth}},\
  }\href@noop {} {\bibfield  {journal} {\bibinfo  {journal} {J. Phys. Chem. B}\
  }\textbf {\bibinfo {volume} {122}},\ \bibinfo {pages} {3296} (\bibinfo {year}
  {2018})}\BibitemShut {NoStop}%
\bibitem [{\citenamefont {Berendsen}, \citenamefont {Grigera},\ and\
  \citenamefont {Straatsma}(1987)}]{Berendsen:87}%
  \BibitemOpen
  \bibfield  {author} {\bibinfo {author} {\bibfnamefont {H.~J.~C.}\
  \bibnamefont {Berendsen}}, \bibinfo {author} {\bibfnamefont {J.~R.}\
  \bibnamefont {Grigera}}, \ and\ \bibinfo {author} {\bibfnamefont {T.~P.}\
  \bibnamefont {Straatsma}},\ }\href@noop {} {\bibfield  {journal} {\bibinfo
  {journal} {J. Phys. Chem.}\ }\textbf {\bibinfo {volume} {91}},\ \bibinfo
  {pages} {6269} (\bibinfo {year} {1987})}\BibitemShut {NoStop}%
\bibitem [{\citenamefont {Phillips}\ \emph {et~al.}(2005)\citenamefont
  {Phillips}, \citenamefont {Braun}, \citenamefont {Wang}, \citenamefont
  {Gumbart}, \citenamefont {Tajkhorshid}, \citenamefont {Villa}, \citenamefont
  {Chipot}, \citenamefont {Skeel}, \citenamefont {Kal{\'e}},\ and\
  \citenamefont {Schulten}}]{Phillips:2005qv}%
  \BibitemOpen
  \bibfield  {author} {\bibinfo {author} {\bibfnamefont {J.~C.}\ \bibnamefont
  {Phillips}}, \bibinfo {author} {\bibfnamefont {R.}~\bibnamefont {Braun}},
  \bibinfo {author} {\bibfnamefont {W.}~\bibnamefont {Wang}}, \bibinfo {author}
  {\bibfnamefont {J.}~\bibnamefont {Gumbart}}, \bibinfo {author} {\bibfnamefont
  {E.}~\bibnamefont {Tajkhorshid}}, \bibinfo {author} {\bibfnamefont
  {E.}~\bibnamefont {Villa}}, \bibinfo {author} {\bibfnamefont
  {C.}~\bibnamefont {Chipot}}, \bibinfo {author} {\bibfnamefont {R.~D.}\
  \bibnamefont {Skeel}}, \bibinfo {author} {\bibfnamefont {L.}~\bibnamefont
  {Kal{\'e}}}, \ and\ \bibinfo {author} {\bibfnamefont {K.}~\bibnamefont
  {Schulten}},\ }\href {\doibase 10.1002/jcc.20289} {\bibfield  {journal}
  {\bibinfo  {journal} {J. Comput. Chem.}\ }\textbf {\bibinfo {volume} {26}},\
  \bibinfo {pages} {1781} (\bibinfo {year} {2005})}\BibitemShut {NoStop}%
\bibitem [{\citenamefont {Pines}(1999)}]{Pines:99}%
  \BibitemOpen
  \bibfield  {author} {\bibinfo {author} {\bibfnamefont {D.}~\bibnamefont
  {Pines}},\ }\href@noop {} {\emph {\bibinfo {title} {Elementary excitations in
  solids}}}\ (\bibinfo  {publisher} {Perseus Books},\ \bibinfo {year}
  {1999})\BibitemShut {NoStop}%
\bibitem [{\citenamefont {Martin}\ and\ \citenamefont
  {Matyushov}(2017)}]{DMscirep:17}%
  \BibitemOpen
  \bibfield  {author} {\bibinfo {author} {\bibfnamefont {D.~R.}\ \bibnamefont
  {Martin}}\ and\ \bibinfo {author} {\bibfnamefont {D.~V.}\ \bibnamefont
  {Matyushov}},\ }\href@noop {} {\bibfield  {journal} {\bibinfo  {journal}
  {Sci. Rep.}\ }\textbf {\bibinfo {volume} {7}},\ \bibinfo {pages} {5495}
  (\bibinfo {year} {2017})}\BibitemShut {NoStop}%
\end{thebibliography}
\end{document}